\documentclass[a4paper,12pt]{article}
\begin{document}
\title{An Epistemological Derivation of Quantum Logic}
\author{John Foy}
\maketitle
\pagenumbering{arabic}

\begin{abstract} 

\vskip .25in
This paper deals with the foundations of quantum mechanics. We start by outlining 
the characterisation, due to Birkhoff and Von Neumann, of the logical
structures of the theories of classical
physics  and 
quantum mechanics, as boolean and modular lattices respectively.
We then derive these descriptions from what we claim
 are basic properties of any
 physical theory 
- i.e. the notion that a quantity in such a theory may be analysed into parts
and that the results of this analysis may be treated in  languages with an underlying
boolean structure. We shall see that in the course of  constructing a model of  a
theory with these properties different
indistinguishable possibilities will arise
for how the elements of the model may be named, that is to say different possibilities
arise
for how they can be associated  with points from Set. Taking a particular collection of
 possibilities gives
the usual boolean lattice of the propositions of classical physics. Taking all
possibilities -  in a sense, the set of all things that may be described by physical
 theories  -
gives the lattice of quantum mechanical propositions. This gives an interpretation of 
 quantum mechanics as the
complete set of such possible descriptions, the complete physical description of the
world.

\end {abstract}

\section{ Introduction}

In physics,  the theory of statics comprises propositions about two basic observables, position and
momentum. These propositions define ranges of values for each observable.

We may then add to this  a notion of phase space and a law of propagation associated with
the physical system.

In a classical physical theory the subsets of phase space are correlated with
propositions about the ranges of values so that there is an obvious correspondence
between the set theoretic operators of union and intersection and the logical
connectives `and' and `or' - with set theoretic inclusion corresponding to logical
implication.

In a quantum mechanical system the propositions concerning ranges of the
observables are correlated with the subspaces of a Hilbert space and the logical
connectives between propositions correspond to set products, sums and complements
with, once again, inclusion corresponding to implication.

Obviously therefore the quantum mechanical and classical propositional calculi differ 
from an
algebraic standpoint. This difference is precisely the following. In the
classical propositional calculus the logical connectives between propositions obey the
distributive law,
\[a\wedge (b\vee c)=(a\wedge b)\vee(a\wedge c)\]
while in the propositional calculus of quantum mechanics $\wedge$ and $\vee$ do not
obey this law although they obey a modified version of it - the weaker modular law,
\[\mbox{ if }a\leq c \mbox{ then } a\wedge (b\vee c)=(a\wedge b)\vee(a\wedge c)\]
In both cases  $\wedge$ and $\vee$ obey the other usual assumptions sufficient to make
$``a\leq b \mbox{ iff } a\wedge b = a \mbox{ (or equivalently } a\vee b =b
\mbox{) "} $ a partial order.

If we assume in addition to the modular law that every proposition can be written as
the union of basic elements -- atomicity -- and any pair of these basic elements have a
common complement -- perspectivity --
then we get the characterisation due to Birkhoff and Von Neumann [1] of the lattice of
quantum mechanical propositions as an infinite, modular, atomic, perspective lattice.

Since all the nonclassical results of quantum mechanics arise from this distinction in
algebraic structure many attempts have been made to explain the need for this modular
law.

Here it is derived from what we take to be the basic requirements demanded of
 any physical theory --- that the theory contains  an operator representing the
 analysis of  quantities into distinct named parts 
--- and that we can treat this operator in  boolean languages i.e. we can identify
 substructures produced by the analysis operator with  sublattices of boolean
lattices subject
to certain consistency requirements.

To produce a model points from some model of set theory, $ Set$,  must be assigned to
 the variables of the theory.
We have not specified which points in particular must be chosen from $Set$ and certain
indistinguishable possibilities arise for models generated in this way.
Choosing a single set of possibilities gives the usual boolean lattice as we would
expect but we show that the quantum 
 mechanical lattice is the lattice generated by taking all consistent possibilities.
 Quantum
mechanics is therefore in a sense the fullest description of the physical world if the world is
 restricted
to what can be modelled by structures of the above kind.

\section{ Definitions }

We have stated that the propositions of classical physics are points in boolean
 lattices. How is such
a lattice defined ?

A structure will be called a lattice iff to any pair of its elements $x$ and $y$
there correspond elements $x\wedge y$ , $x\vee y$ with the operators $\wedge , \vee 
$ (`join' and `meet') satisfying 

\vskip .15 in
\begin{eqnarray}
 x\wedge x = x,\quad x\vee x = x, &\mbox{ (idempotency)}&\\
 x\wedge y = y\wedge x ,\quad x\vee y = y\vee x, &\mbox{ (commutativity)}&\\
 x\wedge (y\wedge z)=(x\wedge y)\wedge z ,\quad  x\vee (y\vee z)=(x\vee y)\vee z,
 &\mbox{(associativity)}&\\
 x\wedge (x\wedge y)= x\vee (x\vee y)=x, &\mbox{ (absorbtion)}&
\end{eqnarray}

 \noindent $ \wedge , \vee $ then also define a partial order given by $x \leq y \mbox{ if }
 x\wedge y = x $ (or, equivalently, $ x\vee y =y) $.

\noindent If in addition  $ \wedge , \vee $ satisfy 
\begin{equation}
 x\wedge (y\vee z)= (x\wedge y)\vee (x\wedge y) 
\end{equation}
the lattice is said to be distributive.

\noindent If instead they satisfy the weaker
\begin{equation}
x\leq z \Rightarrow x\wedge (y\vee z)= (x\wedge y)\vee (x\wedge y) 
\end{equation}
the lattice is said to be modular.

The following is an equivalent formulation of the modular law for finite lattices.
Let $B$ be a lattice and $h$ a positive function, the height function, defined on $B$
where for all $a, b, c,$  in $B$ 
\begin{equation}
h(a \wedge b) = h(a) + h(b) + h(a \vee b)  
\end{equation}
 Then
$B$ satisfies the modular law. Clearly a distributive lattice supports a function
satisfying (7).

\noindent If there exist elements $0,1$ in the lattice satisfying 
\[ 
0 \leq x \leq 1 \quad \forall x
\]
and if for every element $x$ in the lattice there is a $y$ with 
\[ 
x \wedge y = 0,\quad x \vee y = 1
\]
the lattice is said to be complemented.

\noindent A complemented distributive lattice is called a boolean lattice.

\noindent For $x,\ y$ elements of  a lattice, if $\exists z$ with
\begin{eqnarray*}
 x\wedge z = 0, \ x \vee z = 1  \\ 
 y\wedge z = 0, \ y \vee z = 1  
\end{eqnarray*}
then $x$ and $y$ are said to have a common complement $z$.
Such $x$ and $y$ are said to be perspective. If every pair of points in a lattice are
perspective the lattice is said to be perspective.

\noindent Finally, an atom is a point $x$ such that $\forall  y$ in the lattice
\[ y < x \Rightarrow y = 0 \]

\section{ Outline of the Argument}

If we want to produce an analytic description of the world then, formally speaking,
it must at least satisfy the conditions sketched at the end of section 1 above, which we
restate here.
Our central premise is that anything we think of as a physical theory must incorporate
the idea of analysis of individuals into parts and the treatment as far as is
consistently possible of these parts in languages containing (boolean) joins and
meets and that any attempt to construct a model of such a theory will
naturally reflect this.
We can think of the physical theory as containing the 
 collection of statements arising from analysing the
universe into parts through measurement and the further statements that can be made about these parts in
the boolean languages that underly any formal discussion.

We first give a lengthier informal description of the
properties that should be satisfied by   a  structure representing a  physical theory 
and later define them
 precisely.

\vskip .15in
{\bf Firstly,} the structure should be nonempty -- it should contain two points chosen
from set theory call them
 0 and 1, and, further, in the ordering relation we will define on the structure, 
 0 and 1 satisfy  $0 \leq x \leq 1 \ \forall x$  
in the structure.

\vskip .15in
{\bf Secondly,} a  structure representing a physical theory should support an
 operator representing the 
analysis or division of
an element into parts.
Given any element $a$ in the  structure the  structure should also contain an element 
$b_a$  distinct from $a$, with $a = b_a \vee a$
 contained in the  structure and given such a $b_a$ there should also exist a $c_a$ 
in the  structure 
 with $a = b_a \vee c_a$ and such that $\forall e \  c_a \vee e = c_a$
 and $b_a \vee e = b_a$ iff $e=0$.  We claim that to represent the natural notion of
analysis of a whole into parts 
$\vee $ should be  a binary operator generating a partial order. 

It is important to note here that in terms of using this property to generate such a
structure  the elements from $Set$ represented by $ b_a$ and 
$ c_a,$  above are not
uniquely specified --- there are many possible candidates in $Set$ for such elements.

A chain between $a_1$ and $a_h$ in the  structure is a set of distinct points $\lbrace 
a_1,
\dots, a_h \rbrace$ with $ \forall i \ a_i \vee a_{i+1} = a_i  $. 
We claim that to represent  the idea of analysis $\vee$ should satisfy a further
condition. Let a refinement of a chain be a
larger chain containing it.
Then refinements exist subject to the following. For a given $a, b$ in the  structure
either any
path between $a$ and $b$ may be refined or there exists a bound $d(a,b)$ such that
only those
paths of length $<d(a,b)$ may be refined. 

\vskip .15in
We claim that if in addition $\vee$ is 
 consistent
with the existence of a   lattice extending it and satisfying the same properties then
this operator represents our 
natural notion of  analysis  i.e. the operator represents the `part of'
relation arising from measurement in a physical theory. 

\vskip .15in
{\bf Thirdly,} these two principles give rise to many distinct elements and statements
of the form $a = b \vee c$ where $a$, $b$ and $c$ represent points from $Set$. 
Let $M$ be a set of statements generated in this way.
While $\vee$ is consistent with the existence of a lattice with the given properties it does not
generate it. In some cases it may be possible to 
embed $M$ in a natural language in such a way that the boolean operators of
the language generate the lattice. However it may also be that the lattice we require
may not be embeddable in such a language. This is the source of the difficulties in the
interpretation of quantum mechanics. All such difficulties can be reduced to the problem of trying to
construct the above lattice  in terms of an analysis based on what is constructible in
 boolean lattices. The purpose of this paper is to produce such an analysis.

As we have just said it may be possible to make statements about $M$ in a natural language, that is to say
it may be possible to embed $M$ in a boolean lattice in such a way that the combined
structure still satisfies the above requirements on an analytic operator.
A physical theory should include the expansion of the analytic operator by statements
which can be built up by a treatment of the operator in a natural language subject to
a general demand of consistency.

So  a physical theory can make statements about the analytic operator in
a language, L, at least strong enough to contain the statements that could arise in any
natural discussion, i.e. L contains the boolean operators 
 $\wedge_L $,  $\vee_L$ and $'_L$ or complement.
 Consider the 
 statements which we may be able to make about
$M$ in  such a language. 
It may be that there is a 
structure generated from $M$ using  $\wedge_L $, $\vee_L$ and $'_L$ extending $M$ to a boolean
lattice $B_M$
in such a way that  $B_M$ agrees with the lattice operators, height
function and possibly the $0,1$ of $M$ where they overlap
 and --
restricting ourselves to extending the structure as a partition operator -- such that
 the combined
 structure
is  still  capable of extension to a lattice satisfying the properties described above.

One obvious requirement that this extension principle be a consistent one is that we
allow an extension only when it is consistent with all such extensions of the
substructures of $M$.
For a given  $M$ there may be a number of 
 ways of forming  boolean extensions of the substructures of  $M$, call them $M_i$,
 in such a way that they are mutually consistent and consistent with the other
conditions outlined above. 
A collection $\lbrace B_{M_i} \rbrace $ of such extensions may be maximal
i.e. there is no additional substructure $M_j $ such that any $ B_{M_j} $ is consistent
with the collection $\lbrace B_{M_i} \rbrace $. Call such a collection a
cover of $M$. 
 Then the existence of any such cover will  imply the
 existence of a boolean extension of  $M$, $B_{M}$, iff every cover of $M$ contains such
a $B_M$.
The third property that should be enjoyed by a physical theory is  that if this condition holds the 
structure may be extended
by some $B_M$. We are allowing $M$ and its subsets to be extended by the boolean operators
of the language only if this happens in a consistent way.

 We will show that  these properties are  realised in a boolean lattice and any
structure given by these properties contains such a lattice.

\vskip .15in
In describing these properties we alluded to the fact that points chosen to
name variables in the theory are not uniquely specified. It turns out we can generate
 a structure
realising all of these possible choices simultaneously by first extending  the second
property to state that in
addition to $b_a, c_a $ with $b_a \vee c_a = a $ there is also a point $b'_a \neq 
b_a, c_a$ with $b'_a \vee d_a = a $.

 The structure which is generated by extending
Principle II in this way  is  $P_{n-1}$, Birkhoff and Von
Neumann's characterisation of the quantum mechanical lattice.

\section{Construction Principles}

Let $Const = \lbrace c_1, \dots \rbrace $ where $c_i$ are distinct points chosen from 
some model of set theory, $Set$. Equivalently we may define the $c_i$ to be distinct
 elements
in the language and $\forall c_i, c_j, \ c_i \neq c_j $ is true.

In the last section we sketched a set of properties which  a
structure representing a set of physical propositions should have. We shall see that
 these
properties are sufficient to represent such a set i.e. if we recast them
as construction principles the structure they describe is that of the lattice of
physical propositions. We describe these principles below.

\vskip .15in
\textsc{ I } \quad  The structure  should be nonempty.

\textsc{ Principle I } \quad \emph {The  structure should
   contain a substructure consisting of 
 two points chosen from $Const$. Call the points $0$ and $1$. We define $0 \vee 1 = 1$
and in all that follows it is consistent that \ $0\vee x =x, \quad x\vee 1 =1$ for any $x$ we construct. }

\vskip .15in

\textsc{ II } \quad From the principle of analysis of a whole into parts we get
 the following 

\textsc{ Principle II } \quad \emph {
Given any element $a$, in the  structure the  structure should also contain an
 element 
$b_a$,  distinct from $a$ chosen from $Const$, with the statement $a = b_a \vee a$
 contained in the  structure and given such an $a$ and  $b_a$ the structure contains
 a $c_a$ from $Const$ 
 with $a = b_a \vee c_a$ and  such that $\forall e \  c_a \vee e = c_a$
 and $b_a \vee e = b_a$ iff $e=0$.}
 
Whether $b_a, c_a$ are  points already chosen in the structure is not defined.
It is important to note here that in extending the structure according to Principle II
 we have  not specified  which elements of
$Const$ are represented by  $b_a$ and $c_a$. All 
that is demanded of them  is that they should be distinct and different from $a$.
 Principle II 
states the existence of a condition to be satisfied by  points from $Const$ without
actually specifying those points.

We restate here that a chain between $a_1$ and $a_h$ in the theory is a set of distinct points $\lbrace a_1,
\dots, a_h \rbrace$ with $a_i \vee a_{i+1} = a_i \quad \forall i $. 
 Let a refinement of a chain be a
larger chain containing it.

Then Principle II extends the structure subject to
refinements satisfying the following condition; for a given $a, b$ in the structure
either any
path between $a$ and $b$ may be refined or there exists a bound $d(a,b)$ such that any
path of length $<d(a,b)$ may be refined.
Further $\vee $ should satisfy the requirements on a partial order and be consistent
with the existence of a lattice extending it and satisfying the  requirements outlined
above, i.e. 
\indent .15in  (i) Principle II holds in the lattice subject to Principle IV.
\indent .15in  (ii) Refinements exist in the lattice subject to the condition that for
any $a$ and $b$ in the lattice there exists a bound $d(a,b)$ such that just those
paths of length  $\leq   d(a,b)$ may be refined.

\vskip .15in
\textsc{III} \quad
Any natural language in which we might treat the analytic operator in Principle II
contains the boolean operators. We  extend the notion of a physical proposition to
include statements we can make about this analytic principle in a natural language.
Arising from this assumption
the final  principle  is that the structure should include statements which are given
 by the possibility of
extending   substructures 
using the  boolean operators of such  languages. 
Let $M$ be such a substructure  and let  $\vee_M$ and $\wedge_M$ as defined in $M$
be
consistent with the boolean property  (5). Our natural notion of being able
to treat the propositions of $M$ in a rational language amounts to saying that they
may be embedded in a
structure equipped with  boolean operators,  $\vee_L$, $\wedge_L$ and $'_L$, which are 
consistent
with and  extend the  
$\vee$ and $\wedge$ of the original structure in such a way that  $\vee$ and $\wedge$
are consistent with the requirements described in the definition of Principle II.  
Thus the third
 principle generating the structure  is  defined as follows;

  A substructure $M$  comprises a set of statements of the form $a \vee b = c , \
a' \wedge b' = c' ...$ and statements regarding the function $d$, defined above,
 applied to points
in $M, d(a,b) = d_{ab} \dots $.

Let $\lbrace M_i, i \in I \rbrace$ be the set of substructures of $M$. For a given $M_i$
it may be possible to define a structure  $B_{M_i}$, a boolean lattice,
 on points chosen from $Const$,
containing $M_i$ and  consistent with 
the structure as
defined so far (where by consistency we mean that  if we define a function $h$ on the
structure by $d(0,a)= h(a)$, and $h_B$ is a height function on the distributive
$B_{M_i}$, then 
 $\wedge_B $, $\vee_B $, $h_B$, $0_B$ and $ 1_B$ agree with  $\wedge, \vee$, $ h $,
0 and 1, where
defined in the structure so far, and the structure, extended by $B_{M_i}$, can still
be extended to a lattice satisfying the requirements given above ).

Next we define a cover of $M$.
Given a collection of substructures of $M$, $\lbrace M_i : i \in I \rbrace$,
suppose that for each $M_i$ in the collection there exists such a boolean lattice and
that these lattices are consistent in the sense described above with each other and
the rest of the structure so far defined. 
If  $\lbrace B_{M_i} : i \in I \rbrace$ is not contained in a larger collection of
boolean lattices with this property  $\lbrace B_{M_i} : i \in I', I' \supset I \rbrace$
we call $\lbrace B_{M_i} : i \in I \rbrace$ a cover of $M$.
 Such a cover represents a fullest 
possible
mutually consistent treatment of the parts of $M$ in boolean languages in the manner
described.

If the existence of any such  way of consistently extending  parts of $M$ to such
lattices necessarily implies the
existence of some $B_{M}$, i.e. if every cover of $M$ contains a $B_M$,
 then we demand that the structure should be extended by some such  $B_{M}$. In other
 words if
granting that we can treat as many parts as possible of $M$ in a rational language
implies that we have a boolean lattice containing $M$ then we may
add some such  lattice to the structure.  
Recasting this as a construction principle we can say that the structure should
contain that substructure common to all boolean lattices containing $M$ and satisfying
the consistency requirements described above.

\vskip .15in

\textsc{ Principle III } \quad \emph {For $M$ a given  substructure, if all
maximal coverings, $\lbrace B_{M_i}: i \in I, M_i \subset M \rbrace$, contain a
 $B_{M}$, then  the structure should contain a substructure common to all such $B_M$.}

\vskip .15in

We claim that this completely describes our natural notion of what can be said about
the products of analysis in every boolean language.

\vskip .15in
In addition to these three principles we introduce an ad hoc assumption limiting the
depth of the structure .

In their characterisation of the lattice of quantum mechanical propositions 
 Birkhoff and Von Neumann, for the sake of simplifying the proof, restrict their attention
 to lattices of depth bounded by some
$n \in \textbf{N} $ - that is to say  the length of every chain in the lattice is
bounded by $n$. We do not need a restriction on the length of chains in the structures
to prove the general result of this paper characterising the propositional calculi
of the theories of  classical and quantum mechanics. However, in the interest of simplifying
our proof  we too will assume this  ad hoc bound
 on the length of chains in the structure and under this assumption derive from our
generating principles Birkhoff
and Von Neumann's restricted models of the propositional calculi.
Again the structures we get in the absence of this ad hoc assumption are equivalent to
the infinite models of Birkhoff and Von Neumann but the proof is more elaborate.
We introduce the following

\vskip .15in

\textsc{ Ad Hoc Principle IV} \quad
\emph{For some $n \in$ \textbf{N} \  Principles II and III hold subject to the
requirement that
for any $a$ in the structure $d(0,a) \leq n$   }

\vskip .15in
Define $d(0,a) = h(a)$, called the height of $a$.
\vskip .15in

\section{ These four  principles generate and are realised in  a boolean
 lattice.}

 We now show that these four principles  are realised in a boolean lattice of depth $n$ and any
 structure generated by these four
 principles contains such a
boolean lattice. Which is what  we would expect to get from treating the
 products
 of  an analytic principle like II in a language with an underlying boolean operator.

Let $B$ be a model of a boolean lattice of depth $ n$ i.e. $h(1)= n$.
We first show that $B$  realises any
structure constructed by I-IV, and hence these principles are consistent.

\vskip .15in

Principle I stating the existence of a 0 and 1 in $B$ with $0 < x < 1 \ \forall x \in
B$ is obviously satisfied in $B$.

\vskip .15in
Let $C$ be a structure generated by  Principles I-IV and let
$C$ be realised in $B$ i.e. 
there exists a homomorphism $ f$  mapping $\lbrace C \rbrace $ into $B$ and 
$\forall a, b, c \in C \ a \wedge b = c \Rightarrow f(a)\wedge f(b) = f(c),
 \quad a \vee b = c \Rightarrow f(a) \vee f(b) = f(c),
\quad h_{C}(a) = h_B (f(a))$.
Then we will show that the extension of $C$ by an application of Principle II is
realised in $B$.
Let $a$ be a point in $C$ realised in  $B$. Then Principle II states that there
should exist $b_a$ and
$c_a$ with $b_a \vee c_a = a$ and  such that $\forall e \  c_a \vee e = c_a$
 and $b_a \vee e = b_a$ iff $e=0$
subject to this being consistent with the ad hoc
Principle IV, i.e. subject to $h(a) \geq 2 $. But if
   $h(a) \geq 2 $ in $B$  then
obviously $B$ contains such a  $b_a$ and $c_a$ with $b_a \wedge c_a = 0$.
 Hence Principle II is satisfied in $B$.

\vskip .15in

Next we show that an extension of $C$ by an application of Principle III is realised
in $B$.
Let $M$ be a substructure of $C$. $M$ is
realised in $B$. Then Principle III asserts the existence of a substructure
 containing $M$ common to all $B_{M}$. But $B$ itself obviously  realises  such a
 substructure.

\vskip .15in

Since the extensions of $C$ above were subject to the restrictions of Principle IV we
are done. 

\vskip .2in

We now show that  Principles I-IV  generate a model of $B$, the
boolean lattice of depth $n$, and hence that any model generated by I-IV 
contains a submodel   equivalent to
$B$.

Let 1 be given by Principle I. 

By repeated application of Principle II we can construct a tree $T$  of depth $n$ with
$2^n$ points of height 1 (or atoms). Call them $p_1, \dots, p_{2^n}$.
A set of atoms $q_1, \dots, q_{n},$ are said to be independent if for any $q_i$ and
any other $q_{j_1}, \dots, q_{j_k}$ in the set $ q_i \not< q_{j_1}\vee \dots, \vee q_{j_k} $.
 $\lbrace p_1, \dots, p_{2^n} \rbrace $ contains $n$  points
 $q_1, \dots, q_{n}$ generating a boolean lattice of height $n$. We first demonstrate
that   $\lbrace p_1, \dots, p_{2^n} \rbrace $ contains $n$ points $q_1, \dots, q_{n}$
such that $h(q_1\vee \dots \vee q_{n}) =n$.

We show this by induction.  
Set $q_1 =$ any $p_i \in \lbrace p_1, \dots, p_{2^n} \rbrace $. Then $h(q_1) =1$.

We prove the  induction step as follows. We assume there exists a $P_i $ in $T$ 
 with
$h(P_i) =i$ and $q_1, \dots, q_{i}$ atoms in $T$ with $q_j < P_i\ \forall j \leq i$.
By construction of $T$ there is a $P_{i+1}$ in $T$ with $P_{i+1} > P_i$ and a $q_{i+1}$,
 equal
to some $p_j$ , with $q_{i+1} <P_{i+1}$ , $q_{i+1} \neq q_{j} \ \forall j \leq
i$,
and $q_{i+1} \not< P_{i}$ (if we can't find such a $q_{i+1}$ then any $p_j < P_{i+1}$ 
would also be $< P_{i}$ and,
since it follows from the construction of $T$ that $P_{i+1}$ is the join of some set
of $p_j < P_{i+1}$ we would have $ P_{i+1} \leq P_i $ contradicting the fact that the
$P_i$ arose from an application of Principle II to $P_{i+1}$).

Now $q_{i+1} \not< \bigvee_{j \leq i} q_j $ and $h(  \bigvee_{j \leq i+1} q_j) >
 h( \bigvee_{j \leq i} q_j)$. But $h( \bigvee_{j \leq i+1} q_j) \leq
 h( \bigvee_{j \leq i} q_j) + h(q_j)$. Therefore  $h( \bigvee_{j \leq i+1} q_j) =
 h( \bigvee_{j \leq i} q_j) + 1$.

A set of atoms $q_1, \dots, q_{n},$ are said to be independent if for any $q_i$ and
any other $q_{j_1}, \dots, q_{j_k}$ in the set $ q_i \not< q_{j_1}\vee \dots,
 \vee q_{j_k}$.

The lattices in any cover of $M=\lbrace q_1, \dots, q_n \rbrace$ are consistent with
the extension of $M$ as a lattice satisfying a height function i.e. a modular lattice.
But any set of points $\lbrace q_1, \dots, q_n \rbrace$ in a modular lattice with $q_i
\wedge q_j =0 \ \forall i,j \leq n$ and $h( \bigvee_{i \leq n} q_i)=n$ are independent
and their join, having height $n$, if it exists is equal to 1. Hence 
 $B_{\lbrace q_1, \dots, q_n \rbrace}$ is
 unique  and  Principle III
ensures that the structure contains this lattice.

\section{ Construction Principle IIa  }

Principles I-IV generate and are realised by  the lattice of  propositions of classical physics.

As we use each principle to enlarge the structure 
 we choose constants  from $Const$, to satisfy the new relation generated by the 
principle. These new constants need only be related to points already chosen in a
manner implied by the relations generated so far and these are the only relations they
must satisfy.
 Therefore there may be many possible
choices of constant to substitute  in a relation generated by a given principle
 and hence many
distinct possibilities for a structure realising Principles
I-IV.

Let us consider a larger structure than one given by Principles I-IV; the structure, 
call it Q, which comprises all such simultaneously possible structures. What meaning may we attach to
 Q? If a structure generated by   Principles I-IV represents a
physical description of the world then Q would contain  all the physical
descriptions that are simultaneously possible based on our treatment of analysis in a boolean language. Q then
 represents the fullest description of the world if the world is restricted
to statements which belong in some model of this conception of a  physical theory. Q is
generated by strengthening Principle II  in the following way.

In the definition of Principle II, for  a given $a$ in the structure we define elements
$b_a$ and $ c_a$,  chosen from $Const$ , with $b_a \vee c_a = a$. To generate $Q$ we
define the stronger Principle IIa which says there also exists $b_a' \in Const$
distinct from $b_a$ and $c_a$ with $b_a' \vee c_a = a$ and  such that
 $\forall e \  c_a \vee e = c_a$
 and $b_a' \vee e = b_a'$ iff $e=0$.

It is only this  principle which must be strengthened to generate $Q$.
 The 0 and 1
of Principle I are, by definition,  unique and no new points are introduced by
 Principle IV. The points introduced by Principle III are either uniquely defined by
$\wedge $ and $ \vee$ from points already in the structure or else generated by taking
complements. However we will show that the Principles I, III, IV and the augmented
Principle IIa generate a structure in which each point has a non unique complement
and so it is not necessary to augment Principle III.

Formally we extend Principle  II  as follows to generate the
structure comprising all possibilities that arise in creating a model in the manner
described above.

\vskip .15in

\textsc{ Principle $II_A$ } \quad
\emph{ Let $a$ be a point in the structure to which we may apply the analytic 
Principle II (subject
to Principle IV )  . Then in addition to $b_a, c_a$ distinct, with $b_a \vee c_a = a$,
there also exists $b'_a$ with $b_a, c_a, b'_a $ all distinct and $b'_a \vee c_a = a$ }

\vskip .15in
Let $Q$ be the structure  generated by   {\sc Principles I, II}a, and III, 
together with the { \sc ad hoc Principle IV} restricting the range of $h$.

\section{ Q generates and is realised in  the lattice of Quantum Mechanical 
propositions}

We now show $Q$ realises and contains $P_{n-1}$, the projective lattice of dimension
 $n-1$,
Birkhoff and Von Neumann's characterisation of the lattice of quantum mechanical
propositions. They define $P_{n-1}$ as the modular atomic perspective 
lattice of
height bounded by $n$ consisting of the subspaces of the projective lattice $P_{n-1}$
under set intersection and linear sum.

 However the following equivalent characterisation will also be
useful.

\vskip .15in

\noindent $P_{n-1}$ is  a lattice defined as follows;

Call the atoms of the lattice 'points', the elements $\lambda$ with   $h ( \lambda ) =2
$ 'lines', and the elements $\pi$ with   $h ( \pi ) =3 $ 'planes'. We say  a
point, $p$, is on a line, $\lambda$,  when $p \leq \lambda$ in the lattice.
For a line, $\lambda $, and  a plane, $\pi$, if $\lambda \leq \pi$ the line is said to
lie in the plane.

$P_{n-1}$ is then defined to have the following properties.

\noindent {\bf P1} \quad Two distinct points are on one and only one line.

\noindent {\bf P2} \quad If two lines lie in the same plane they have a nonempty
intersection. 

\noindent {\bf P3} \quad Every line contains at least three points. 

\noindent {\bf P3} \quad The set of all points is spanned by $n$ points but not by
fewer than $n$ points. i.e. there is a set of points ${p_1, \ldots, p_n}$ such that for
$p$, a point in $P_{n-1},\quad p \leq p_1 \wedge p_2, \ldots, p_n $.

\vskip .15in

\subsection{ $P_{n-1}$ contains $Q$ }

Let $P_{n-1}$ be a modular atomic perspective lattice of height bounded by $n$.
We will show $P_{n-1}$ realises any statements constructed according to Principles I,
 IIa, III
and IV (and hence these Principles are consistent).
\vskip .15in

 $P_{n-1}$ obviously realises points 0 and 1 satisfying Principle I.

\vskip .15in

Let $Q$ , a structure generated by I, IIa, III and IV, be realised in $P_{n-1}$.
Then as in the earlier case, given any $a$ in $Q$ to which Principle IIa
can be applied subject to   Principle IV,
 $a$ is realised as an
element of height $ > 1$ in $P_{n-1}$ and hence $\exists b_a, c_a \in P_{n-1}$ with $b_a
\vee c_a = a$ and $b_a \wedge c_a =0$ as required.
Since $P_{n-1}$ is perspective $\exists b_a \in P_{n-1}, b'_a \neq b_a, c_a $ with $
b'_a \vee c_a = b'_a \vee b_a = a$ and  $b'_a \wedge c_a =0$. Hence  $P_{n-1}$ realises this additional 
extension of $Q$ arising from the 
application of Principle IIa (subject to Principle IV).

\vskip .15in

Let $M$ be a substructure of $Q$ realised in $P_{n-1}$ and such that Principle III
 generates a substructure of a boolean lattice,
$B_M$, containing $M$ and consistent with Principle IV as described above.

Since $M$ is realised in
 $P_{n-1}$
  each point in $M$ may  be
expressed as a join of atoms in  $P_{n-1}$.
Then there is a boolean lattice in  $P_{n-1}$ realising a $B_M$, 
 i.e.  there is a set of independent atoms in $P_{n-1}$,  ${p_1, \dots, p_n}$
 generating a boolean lattice 
in the same relation to $M$ as some $B_M$ .

Suppose  such a set does not exist. Let
$p_1, \dots, p_n$ be a set of atoms 
 in  $P_{n-1}$
generating the points representing $M$. Further there is no smaller set of atoms with
the same property. 
Then by hypothesis $\exists I, J \subset \lbrace 1,\dots ,n \rbrace I \cap J
=\emptyset$ and $\bigvee_{i \in I} q_i \wedge \bigvee_{j \in J} q_j \not= \emptyset$.
Let $B_{M_i}$ be some boolean sublattice of  $P_{n-1}$  containing the $q_i, i \in I$.
Then the existence of a boolean lattice in  $P_{n-1}$ containing $M$ is contradicted
by the existence of the $q_i, i \in I$ with the above properties and therefore we have
 a cover of $M$
that 
contradicts the condition for the application of Principle III.

Hence  $P_{n-1}$ realises  an extension of $Q$ by application of Principle III.

\vskip .15in

 In the above  $P_{n-1}$ was shown to  realise  Principles IIa and III subject to 
Principle IV.

\subsection{  $Q$ contains $P_{n-1}$ }

Here we show that  a structure  generated by  I, IIa, III and IV, 
satisfies conditions (i)-(iv) defining $P_{n-1}$. 
Let $Q$ be such a structure.

(i) For any two distinct atoms $a,b \in Q \  a \vee b \in Q$.

Let $a,b$ be atoms in $Q$. In any given cover of $M=[a,b]$ the lattices in the cover
are consistent with the existence of $a \vee b$ and $a \wedge b$ satisfying a modular
height function i.e. $h(a \vee b) = 2, h(a \wedge b ) = 0$. They are also consistent
with Principle II by which
  $\exists c$ with $c\vee (a \wedge b) =1$ and $\forall e \  c\vee e = c, \
 e\vee (a \wedge b)=(a \wedge b)\  \Rightarrow e=0$. And , as in the earlier case, 
 repeated application of Principle II
to $c$  shows the
 existence of a set of distinct atoms $q_1, \dots, q_{n-2}$ s.t. in a modular lattice 
 $ \lbrace q_1, \dots q_{n-2},a,b \rbrace $ generate
a $B_{\lbrace q_1,\dots,q_{n-2},a,b \rbrace}$ and
 so any cover  contains such a boolean lattice  and $Q$ contains $a \vee b$.

(ii) Every line contains a third point.

Let $a \vee b$ define a line $L, \ a,b,$ atoms in $Q$. Then by Principle IIa there exists
$b'$ in $Q$ with $a \vee b' = L$.

(iii)Two lines $L, L'$ lying in the same plane $P$ have a non empty intersection.

In any given cover of $M=[L_1,L_2, P]$ the lattices in the cover
are consistent with the existence of $L_1 \vee L_2$ and $L_1 \wedge L_2$ with height 
determined as follows. $h(L_1 \vee L_2) > 2 $ since $L_1 \neq L_2$, 
 $h(L_1 \vee L_2) \leq 3 $ since $L_1 < P $, $L_2 < P $ and so $h(L_1 \vee L_2) = 3 $,
$L_1 \vee L_2 = P $ and
$h(L_1 \vee L_2)
=h(L_1) + h(L_2) -  h(L_1 \wedge L_2) = 1.$ 
Set  $L_1 \wedge L_2 = q' $ then any set of lattices in the cover are consistent with
the existence of a pair of points $q_1, q_2$ not contained in any of the other lattices
with 
$h(q_1)= h(q_2) = 1$,  $q_1 \vee L_1 = L_1$,  $q_2 \vee L_2 = L_2 $ and hence since
these obey  a height function $q_1 \vee q' = L_1$,  $q_2 \vee q' = L_2 $,
  $q_1 \vee q_2 \vee q_i = P $. 
Arguing as in (i) we can show the existence of a set of atoms 
$\lbrace q_4, \dots ,q_n \rbrace$ such that any modular lattice  containing 
$\lbrace q_1,q_2, q', q_4, \dots ,q_n \rbrace$ contains a  $B_{ \lbrace L_1, L_2, P
\rbrace}$.

Therefore there exists a $B_{ \lbrace L_1, L_2, P
\rbrace}$ in every
cover containing  $L_1 \wedge L_2$ with  $h(L_1 \wedge L_2) =1$ and so  
$h(L_1 \wedge L_2) =1$ in $Q$.

(iv) Since Principle IIa  implies Principle II  and we have already 
shown Principles
I, II, and III subject to Principle IV generate a boolean lattice of height $n$
condition {\bf P3} is satisfied in $Q$.

\section{ Conclusion }

In this paper we have shown how the exotic modular perspective lattice of quantum
mechanical
propositions $P_{n-1}$ can be constructed  from a more intuitive set of assumptions about a physical
theory - in particular our assumptions about what constitutes an analytic procedure and
those about the logical structure of the language in which we deal
with this analytic procedure. Forming a model of everything we can construct in this way
 we generate the lattice of Quantum Mechanical propositions.

Any physical theory contains  a notion of measurement founded on analysis or the
division of the whole into parts. This idea of `part of' is represented by a partial
ordering relation . If this partition arises from some finite process `inf' and `sup'
may be taken giving a binary relation $\wedge$ which can, potentially, be extended to a
modular complemented lattice.

 If the lattice does not contain perspective elements we can construct this extension
 in actuality as follows. Any natural language in
which we treat the partition operator contains the boolean operators $\wedge_L$ and
$\vee_L$ and $'_L$
which may be used subject to requirements of consistency to extend the  structure generated by the partition operator to a
lattice.

However in attempting to construct a representation of partition  in this way many points
from Set may be chosen to represent the elements of the structure  as they are
generated  - i.e. the elements may be named in a variety of ways.

For a particular set of such choices the partition  structure can be extended  to a boolean lattice
 by the boolean
operators of the language in the manner described above.

If we consider the structure generated by each such set of choices to be a
legitimate set of physical propositions the collection of all simultaneously possible 
 such sets will be the
fullest physical description of the world.

However
the complemented modular lattice to which the partial order can potentially be
expanded now contains perspective elements  and hence can not be
expanded to a  boolean lattice  by the boolean operators of the language. Instead we
 can only embed
fragments of this structure representing partition in our boolean languages. The
model that is implied by the possibility of expanding these fragments in this way
  is
$P_{n-1}$.

The central point to the development of $P_{n-1}$ described here is that the expansion
 in
terms of the usual boolean operators of the natural notion of partition,
when we take in to account the freedom that arises in naming the elements of the
structures, is sufficient 
to generate the problematic modular
operator of $P_{n-1}$.

\vskip .15in
What are the consequences of this development of the quantum mechanical lattice of
 propositions for
the interpretation of  quantum mechanics ?

 Wave particle duality as observed in the two slit experiment is a consequence of 
$P_{n-1}$ satisfying the perspectivity property  derived above. Another consequence of the
independence of the perspectivity relation, given by Principle IIa, from the
 principles generating a boolean
lattice is that there exist propositions in $P_{n-1}$ outside a given boolean lattice in
$P_{n-1}$. This notion of independence may be reformulated to give many of the
interpretations of  Quantum Mechanics. e.g. technically randomness can be defined in 
terms of such
independence and this leads in a natural way to the probablistic interpretation of 
 Quantum Mechanics . 

What has been done here is to provide from an examination of the nature of physical
theories a set of generating principles for physical propositions that are formally
stronger than, or
independent of, principles generating a given boolean lattice of physical propositions
--
such formal independence being already the basis for most interpretations of 
 Quantum Mechanics .

\section{Bibliography}

[1] `The logic of Quantum Mechanics' G.Birkhoff, J. von Neumann {\it Annals of
Mathematics} {\bf 37} 1936, 823-43.

\end{document}